# Parallel-leaky capacitance equivalent circuit model for MgO magnetic tunnel junctions

Ajeesh M. Sahadevan, Kalon Gopinadhan, Charanjit S. Bhatia, and Hyunsoo Yang[a)]

Department of Electrical and Computer Engineering, NUSNNI-Nanocore, National University of Singapore, 4 Engineering Drive 3, Singapore 117576, Singapore

The capacitance of MgO based magnetic tunnel junctions (MTJs) has been observed to be magnetic field dependent. We propose an equivalent circuit for the MTJs with a parallel-leaky capacitance ($C_l$) across the series combination of geometric and interfacial capacitance. The analysis of junctions with different tunneling magnetoresistance values suggests higher $C_l$ for low TMR junctions. Using Cole-Cole plots the capacitive nature of MTJs is manifested. Fitting with Maxwell-Wagner capacitance model validates the *RC* parallel network model for MTJs and the extracted field dependent parameters match with the experimental values.

[a)] e-mail address: eleyang@nus.edu.sg

Magnetic tunnel junctions (MTJs) have played a key role in the development of hard disk drive read heads as well as magnetic random access memories.[1-3] For high speed applications, the product of resistance and capacitance values (*RC* time constant) of MTJs plays an important role. Capacitance occurs when electrical conductors are separated by an insulator and its voltage dependence has been widely used to characterize a variety of physical properties. For example, properties such as the dielectric constant and loss of the insulating layer in metal-insulator-metal structures are revealed by capacitance measurements. In order to characterize the density as well as the distribution of interface trap states at the oxide-semiconductor interface, capacitance-voltage (*C*-*V*) characterization methods have proven to be a very efficient technique.[4]

With extremely thin dielectric materials (~20 Å) in MTJs, the capacitance is largely influenced by the interface properties of the structure as the electric field penetration becomes significant.[5] Since the interfaces are very critical in determining the tunneling magnetoresistance (TMR)[6, 7] which is one of the key attributes of MTJs, the study of capacitance can be a useful tool for a better understanding of MTJs. After theoretical predictions of spin dependent capacitances in ferromagnetic systems[8, 9], the tunneling magnetocapacitance (TMC) effect in $Al_2O_3$ based MTJs has been reported.[10-12] However, the experimental situation for the TMC effect in MgO based MTJs is less clear with one group suggesting the existence of the TMC effect and the other group showing the absence of TMC.[13,14] The concept of negative interfacial capacitance is also controversial which has been introduced to account for the measured capacitance exceeding the geometric capacitance. On the other hand, several reports in $Al_2O_3$ based systems have observed a positive interfacial capacitance.[10, 15, 16]

In this letter, we present the magnetic field dependence of capacitance in MgO MTJs as observed for all our junctions. A parallel-leaky capacitance based equivalent circuit has been proposed in order to account for a larger value of measured capacitance than that of the geometric capacitance. The capacitance and frequency dependent characteristics of high TMR samples are compared with a low TMR sample having a highly leaky tunnel barrier. Using impedance spectroscopy it is shown that the MTJ systems are capacitive and the experimental observations can be explained using an *RC* parallel network. The frequency and voltage dependence of TMR and TMC for the two representative junctions are also investigated.

The MTJs with a structure of 100 Ta/300 $Ir_{22}Mn_{78}$/6 $Co_{40}Fe_{40}B_{20}$/30 $Co_{70}Fe_{30}$/8 Ru/27 $Co_{70}Fe_{30}$/8 Mg/14 MgO/20 $Co_{40}Fe_{40}B_{20}$/50 Ta/50 Ru (all thickness in Å) have been grown using magnetron sputtering in an ultra-high vacuum chamber. The MgO barrier is formed by the reactive sputter deposition of Mg in Ar-$O_2$ plasma (~2% oxygen). The Mg layer prevents the oxidation of the underlying ferromagnetic electrode and is converted to MgO by reactive oxygen introduced into the sputter chamber during the deposition of the MgO layer.[17, 18] The samples are annealed at 300 ˚C for 30 minutes under 1 T magnetic field and then MTJs are fabricated in a current perpendicular-to-plane configuration using a combination of Ar ion-milling and photolithography processes. The capacitance has been measured using an Agilent E4980A Precision LCR meter in the frequency range from 100 Hz to 2 MHz, and the temperature dependence has been studied using a cryostat under high vacuum conditions ($< 1\times10^{-7}$ Torr). Prior to every measurement the effect of stray capacitances and inductances from the measurement probes and co-axial cables is compensated using standard open and short corrections.[19]

A typical TMC is shown in Fig. 1(a) with a plot of the junction resistance with varying magnetic fields. All our MTJs show a negative TMC value smaller than the TMR value as reported previously.[13] The negative TMC ratio indicates that a higher capacitance value for the parallel (P) state than that for the antiparallel (AP) state as predicted by calculations.[20] Although some of the earlier reports suggest interfacial charge accumulation to be independent of the conduction process, a recent report has proposed spin dependent charge accumulation at the interfaces as a reason for the observation of magnetic field dependent capacitance or TMC in MTJs.[12] A simple qualitative way to understand the negative TMC is to consider that the capacitance follows the conductance.[19] For the P configuration the higher value of conductance results in a greater amount of charge accumulation and hence a capacitance value more than that of the AP state. Along with the magnetic field dependence, the voltage or temperature dependence of capacitance has also been observed to be in opposite sense to that of the resistance of the MTJs. It is clear that the TMC effect is correlated with the TMR effect from the fact that the switching fields in Fig. 1(a) for resistance and capacitance match well. However, we did not observe any strong correlation between the TMR and the TMC values. Rather, we found the *RC* time constant is proportional to the TMR ratio as shown in Fig. 1(c).

The *RC* time constant of the MTJs has similar magnetic field dependence to the TMR as shown in Fig. 1(b), suggesting different switching speeds between the P and AP states. Figure 1(c) shows a plot of the relative change of the *RC* time constant for the P and AP states for various samples, defined by

$$TM_{RC} = \frac{(RC)_{AP} - (RC)_{P}}{(RC)_{P}} \quad (1)$$

The value of $TM_{RC}$ is higher for devices with a higher TMR, which indicates a larger asymmetry in the switching speed between the P and AP states for a high TMR sample. The synthetic antiferromagnet part of the structure (CoFe/Ru/CoFe) also contributes to the change of the capacitance as can be seen between 0 and 0.2 T in Fig 1(a). In such a magnetic multilayer without any dielectric, only interfacial charge accumulation can lead to the field dependent capacitance.

The concept of negative interfacial capacitance for the MgO tunnel junctions was introduced in order to explain a larger effective capacitance than the geometrical capacitance.[13, 14] However, the physical origin of negative interfacial capacitance is not clear and several reports in $Al_2O_3$ based MTJs and MIM structures have consistently demonstrated a positive interfacial capacitance value.[10, 15, 16] There have been several reports of negative capacitance in Schottky barriers.[21-23] In these cases, however, at high bias voltage injected high energy carriers knock out charges trapped at interfaces, resulting in the depletion of interfacial charges instead of accumulation similar to impact ionization. This leads to reduction in the measured capacitance with increasing bias voltage, and a negative capacitance value is induced when the bias voltage is sufficiently high. However, there has been no report showing such evidence in MTJs. Furthermore, the observation of a positive capacitance value in pure metallic structures such as pseudo-spin-valves[24] and nanowires[25] is due to the positive interfacial capacitance ($C_i$), in which the geometric capacitance ($C_g$) is absent. Therefore, the origin of a negative interfacial capacitance in MTJs is not clear.

We propose an equivalent circuit for MTJs based on a parallel-leaky capacitance ($C_l$) in parallel to the series combination of $C_g$ and $C_i$ as shown in the inset of Fig. 1(c). The MTJs are in general highly leaky capacitors with very high leakage currents. For metal-oxide-semiconductor structures times domain techniques have been proposed to extract accurate values of capacitance for highly leaky dielectrics.[26-28] Using this equivalent circuit, the total or the measured capacitance ($C_m$) for MTJs is given by

$$C_m = C_l + \frac{C_g C_i}{C_g + C_i} \quad (2)$$

The value of $C_l$ is similar to $C_m$, since the $C_g$ is typically much smaller than $C_m$. The leaky components include $C_l$ and a resistor $R_l$ that promote a net flow of current through the MTJ. Because of our large junction sizes ranging from 70 to 2100 μm$^2$, it is highly possible to have localized spots across the junction area in which there is more accumulation of charges, resulting in multiple effective capacitors connected in parallel. The multiple capacitances are represented by $C_l$ in our model, which contributes most to the measure capacitance value because of parallel connections. More leaky MTJs with higher capacitance values tend to show less TMR values [Fig. 1(d)] and demonstrate a less bias voltage dependence of TMR/TMC as will be discussed later. The higher capacitance can be attributed to the presence of additional leakage paths that reduce the TMR ratio. The above observation provides support that our model describes the MTJs correctly. The basic difference between the conductance (1/resistance) and the capacitance is that the former represents the delocalized states and the latter is more sensitive to the localized states. The observed low TMC values compared to the TMR ratios might suggest the presence of localized trap states in MTJs. The effective capacitor due to these localized states is in parallel to $C_g$ and is represented by $C_l$.

In order to validate our model, we compare the characteristics of a low and a high TMR junction using impedance measurements ($Z = R + jX$). The Cole-Cole diagrams for the MTJs are plotted with the resistance ($R$) and the reactance ($X$) at different frequencies from 5 kHz to 2 MHz as shown in Fig. 2. Both the low and high TMR junctions are capacitive in the P and AP states, as the trajectory lies in the fourth quadrant. The data for the high TMR device can be fitted with a semi-circle as shown in Fig. 2(a) and 2(b), however the fitting using a semi-circular function is not possible for the low TMR device (high $C_l$), indicating deviation from the ideal Maxwellian behavior of dielectric [Fig. 2(c) and 2(d)].[29] The frequency dependence of the real and imaginary parts of the impedance is shown in Fig. 3 for the high TMR [Fig. 3(a)] and the low TMR [Fig. 3(b)] devices. With both the MTJ systems equivalent to the *RC* parallel circuit, their frequency dependence plots can be fitted using the Maxwell-Wagner capacitance model,[24, 25, 30, 31] in which two *RC* networks are connected in series with one part being magnetic field dependent and the other being field independent. The field independent part can be removed by taking the difference of the impedances in the P and the AP states. The change of the impedance between the P and AP states can be represented as $\Delta Z = \Delta R + j\Delta X$ where

$$\Delta R = \left( \frac{R_{AP}}{1+4\pi^2 f^2 R_{AP}{}^2 C_{AP}{}^2} - \frac{R_P}{1+4\pi^2 f^2 R_P{}^2 C_P{}^2} \right) \tag{3}$$

$$\Delta X = -2\pi f \left( \frac{C_{AP} R_{AP}{}^2}{1+4\pi^2 f^2 R_{AP}{}^2 C_{AP}{}^2} - \frac{C_P R_P{}^2}{1+4\pi^2 f^2 R_P{}^2 C_P{}^2} \right). \tag{4}$$

The extracted fitting parameters are very close to the experimentally obtained values as listed in Fig 3(c) and 3(d), which in turn supports the authenticity of the *RC* parallel network for MTJs with low as well as high TMR ratios. The $C_P$ and $C_{AP}$ are equivalent to $C_m$ at different fields, which is a combination of the three capacitors ($C_g$, $C_l$, and $C_i$). Since $C_l$ is close to the $C_m$ value, the extracted $C_P$ and $C_{AP}$ give an idea of the $C_l$ values in the MTJs.

Finally, the frequency and bias voltage dependence are carried out. Figure 4(a) reveals that the various MTJ devices with high TMR values have very little frequency dependence at room temperature as well as at low temperature. However, for a leaky tunnel barrier we observe a sharp drop in the value of TMR at frequencies higher than 300 kHz, as shown in Fig. 4(b) regardless of the temperature. A similar drop in TMR with frequency has been observed in thick tunnel barriers and attributed to capacitive leakage paths that suppress the spin dependent conduction process at higher frequencies.[13] The voltage dependence of the ratio of TMR and TMC is suppressed in the leaky tunnel barrier [Fig. 4(d)] compared to that of a high TMR junction in Fig. 4(c). Another point to note is that the TMC curve overlaps the TMR curve for the low TMR junction in Fig. 4(d), while the TMC is less sensitive to bias voltage than TMR for the high TMR device in Fig. 4(c). Both effects can be attributed to the fact that the low TMR device closely resembles an ohmic-like junction with a very high leakage current compared to an ideal tunnel junction with a high TMR ratio.[18]

In conclusion, our results clearly demonstrate the observation of TMC in MgO based MTJs. An equivalent MTJ circuit is proposed with an additional capacitance in parallel with the series combination of interfacial and geometric capacitances, which can explain the measured capacitance values of different devices with various TMR ratios. Fitting using the Maxwell-Wagner capacitance model also validates the *RC* parallel circuit for the MTJs with extracted parameters close to the experimental values.

This work is supported by the Singapore National Research Foundation under CRP Award No. NRF-CRP 4-2008-06.

FIG. 1. (a) The magnetic field dependence of resistance and capacitance of an MTJ with the junction area of ~70 μ$m^2$. The TMR is 116%, while the TMC is 17% at 1 MHz. The TMR ratio is defined by $(R_{AP}-R_P)/R_P$, where $R_P$ and $R_{AP}$ are the junction resistance in the P and AP alignment, respectively. The TMC ratio is defined by $(C_P-C_{AP})/C_{AP}$, where $C_P$ and $C_{AP}$ are the junction capacitance in the P and AP alignment, respectively. (b) The magnetic field dependence of the *RC* time constant for the same device with a relative difference of 83% between the P and AP states. (c) The dependence of $TM_{RC}$ on the TMR shows greater changes in *RC* time constant for higher TMR devices. The inset shows the equivalent circuit for the MTJ with a parallel leaky capacitor ($C_l$) across the series combination of the geometric capacitance ($C_g$) and interfacial capacitance ($C_i$). (d) The relationship between the capacitance and the TMR. All data are at room temperature.

FIG. 2. Cole-Cole plots for low and high TMR junctions (10 kHz to 2 MHz). For the high TMR (300%) junction with the junction area of ~70 μ$m^2$ in (a) and (b), the semi-circular fits match well with the data, while for the low TMR (8%) junction with the junction area of ~70 μ$m^2$, the data significantly deviates from semicircular fits in (c) and (d). All data are at 20 K.

FIG. 3. The ΔX values for the high TMR (300%) and the low TMR (8%) junctions are shown in (a) and (b), respectively. ΔR is shown in the insets. For both junctions, the fitting parameters are close to the experimental values as list in (c) and (d) below the corresponding figures. All data are at 20 K.

FIG. 4. Frequency dependence of TMR for high TMR junctions (a) and a low TMR device (b). Normalized bias dependence of both TMC and TMR for the high TMR (300%) device (c) and the low TMR (8%) device (d) at 50 kHz and 20 K.

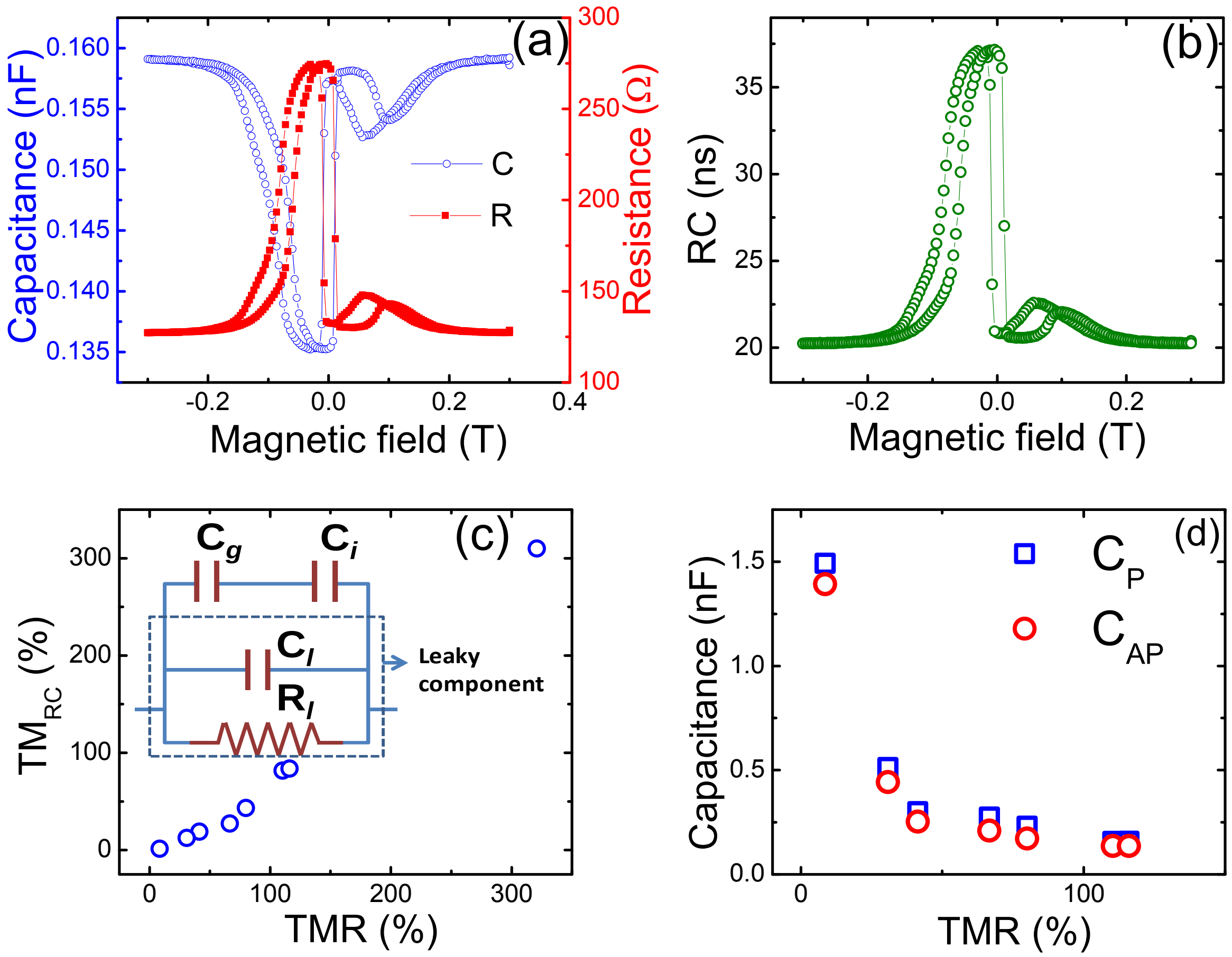

Figure 1

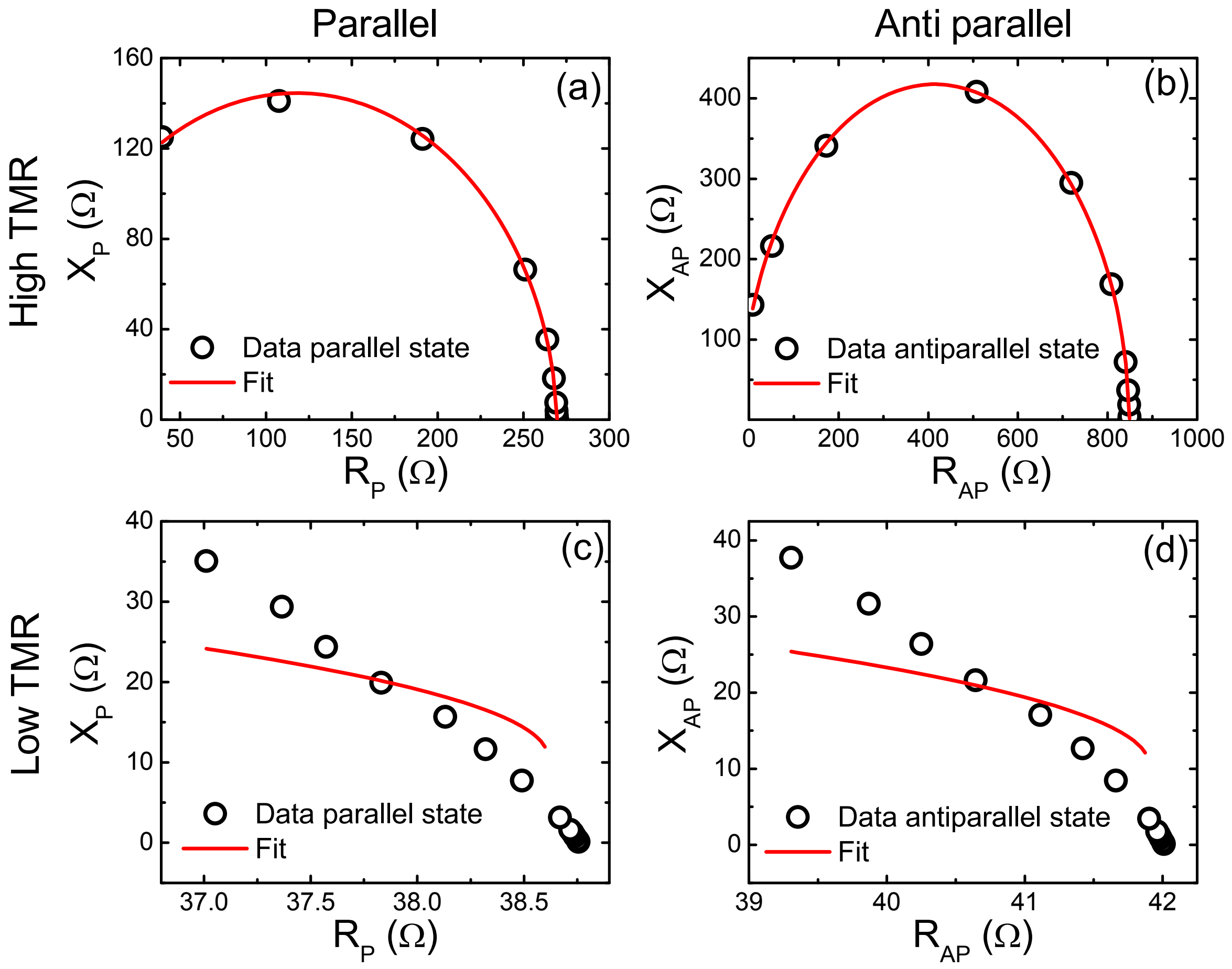

Figure 2

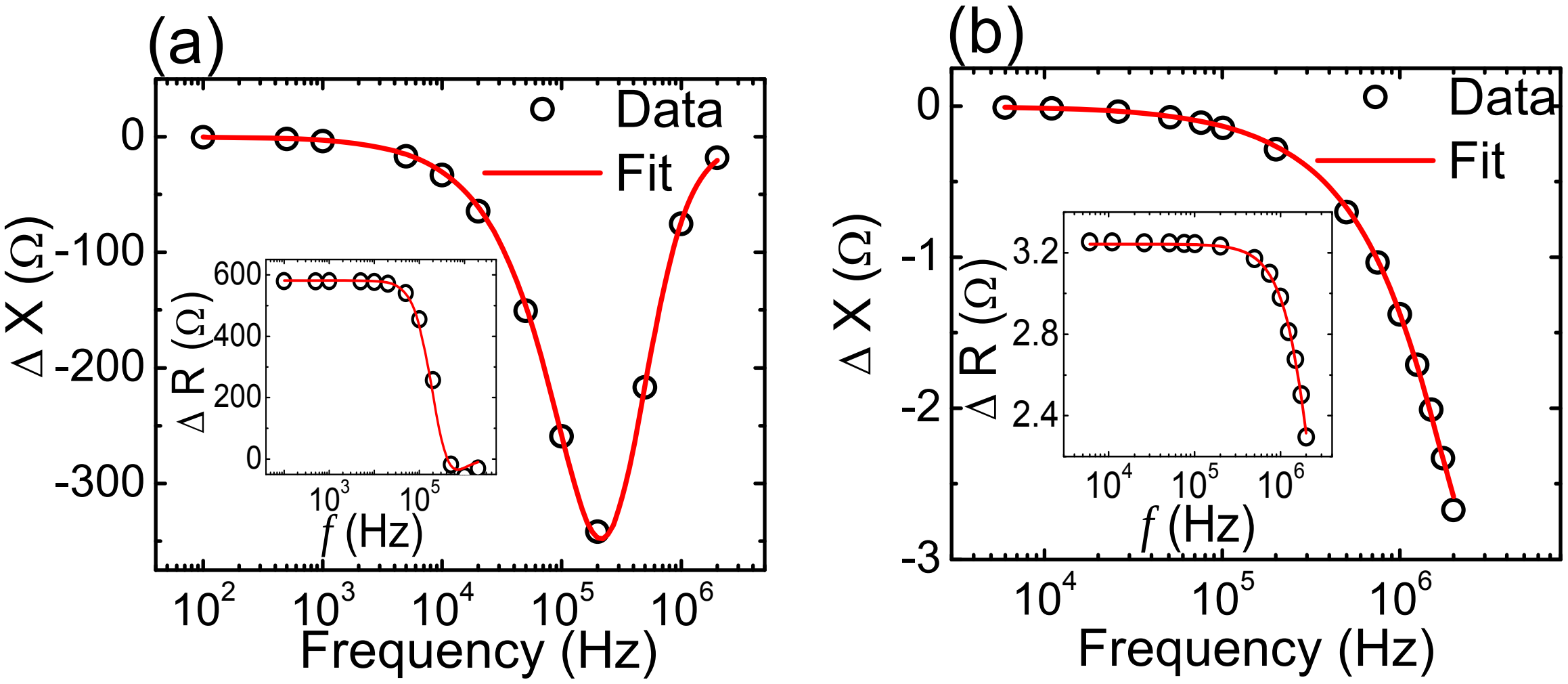

(c)

| | Fitting parameters | Experimental data |
|---|---|---|
| $C_P$ | 0.847 nF | 0.783 nF |
| $C_{AP}$ | 0.757 nF | 0.765 nF |
| $R_P$ | 261.660 Ω | 268.606 Ω |
| $R_{AP}$ | 848.479 Ω | 835.944 Ω |

(d)

| | Fitting parameters | Experimental data |
|---|---|---|
| $C_P$ | 1.671 nF | 1.745 nF |
| $C_{AP}$ | 1.356 nF | 1.622 nF |
| $R_P$ | 37.995 Ω | 38.756 Ω |
| $R_{AP}$ | 43.985 Ω | 42.009 Ω |

Figure 3

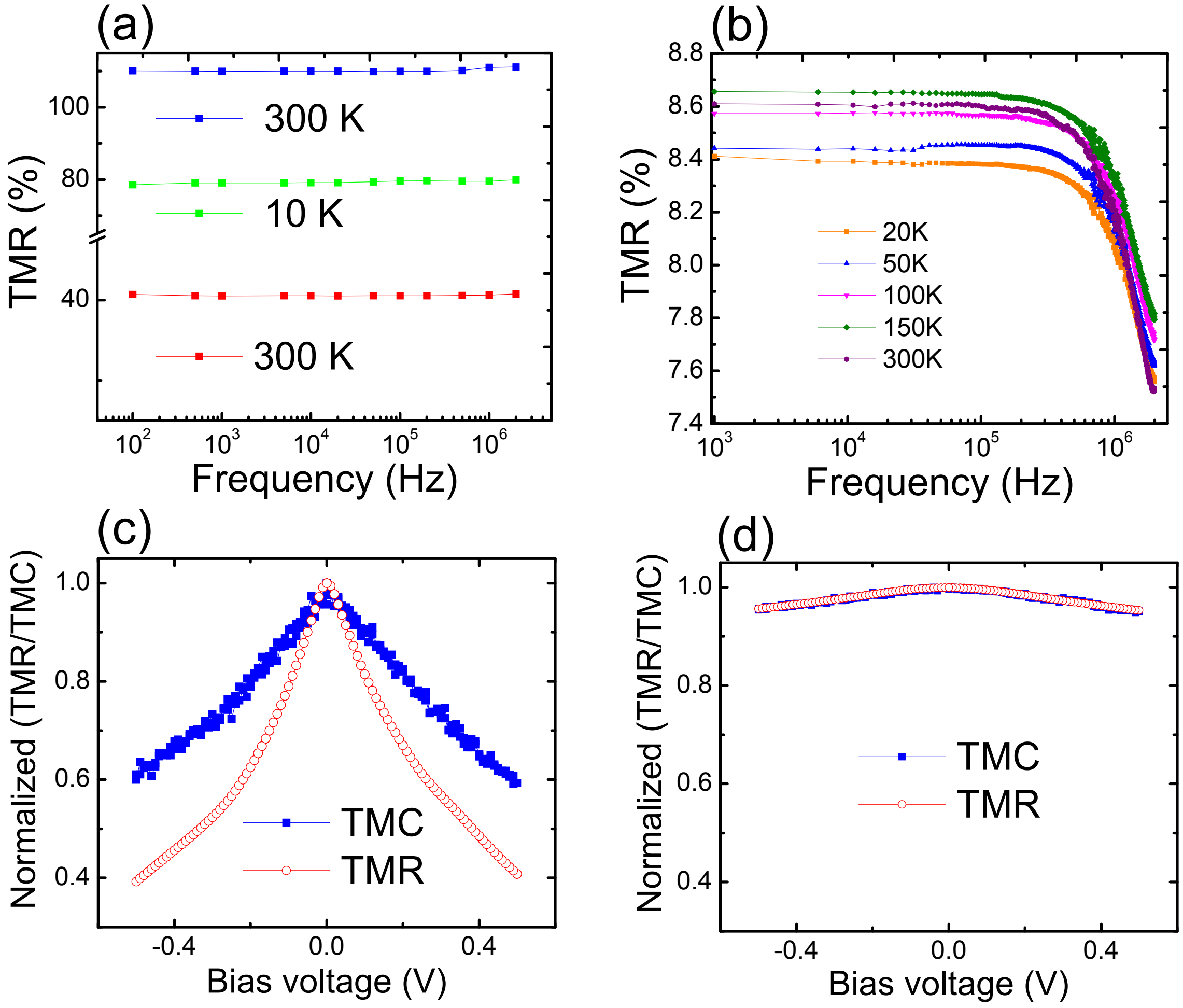

Figure 4